\documentclass[aps, twocolumn,superscriptaddress, 
amsmath,amssymb,reprint,numbers,noeprint, floatfix,longbibliography]{revtex4-1}

\usepackage{graphicx}
\graphicspath{
      {figures/}
      {../figures/}
    }
\usepackage{dcolumn}
\usepackage{bm}

\usepackage{float}

\usepackage{physics}

\usepackage{color}
\definecolor{red}{rgb}{1,0,0}
\definecolor{blue}{rgb}{0,0,1}

\newcommand{\nvn}{\mathrm{NV}^-}
\newcommand{\nvz}{\mathrm{NV}^0}

\newcommand{\sivn}{\mathrm{SiV}^-}
\newcommand{\sivz}{\mathrm{SiV}^0}

\newcommand{\SIsetup}{1}
\newcommand{\SIbeamsize}{2}
\newcommand{\SIrtnv}{3}
\newcommand{\SImodel}{4}
\newcommand{\SIsivn}{5}
\newcommand{\SIsivz}{6}

\begin{document}
\title{Rapid, in-situ neutralization of nitrogen- and silicon-vacancy centers in diamond using above-band-gap optical excitation}

\author{Christian Pederson}%
\thanks{These two authors contributed equally}
\email{cpederso@uw.edu}
\affiliation{University of Washington, Physics Department, Seattle, WA, 98105, USA}%

\author{Nicholas S. Yama}%
\thanks{These two authors contributed equally}
\email{nsyama@uw.edu}
\affiliation{University of Washington, Electrical and Computer Engineering Department, Seattle, WA, 98105, USA}%

\author{Lane Beale}%
\affiliation{University of Washington, Physics Department, Seattle, WA, 98105, USA}%

\author{Matthew L. Markham}%
\affiliation{Element Six, Global Innovation Centre, Fermi Avenue, Harwell Oxford, Didcot, Oxfordshire, OX11 0QR, UK}%

\author{Kai-Mei C. Fu}%
\affiliation{University of Washington, Electrical and Computer Engineering Department, Seattle, WA, 98105, USA}%
\affiliation{University of Washington, Physics Department, Seattle, WA, 98105, USA}%
\affiliation{Physical Sciences Division, Pacific Northwest National Laboratory, Richland, Washington 99352, USA}

\begin{abstract}
    The charge state of a quantum point defect in a solid state host strongly determines its optical and spin characteristics.
    Consequently, techniques for controlling the charge state are required to realize technologies such as quantum networking and sensing.
    In this work we demonstrate the use of deep-ultraviolet (DUV) radiation to dynamically neutralize nitrogen- (NV) and silicon-vacancy (SiV) centers.
    We first examine the conversion between the neutral and negatively charged NV states by correlating the variation of their respective spectra, indicating that more than 99\% of the population of NV centers can be initialized into the neutral charge state.
    We then examine the time dynamics of bleaching and recharging of negatively charged $\sivn$ centers and observe an 80\% reduction in $\sivn$ photoluminescence within a single 100-\textmu s DUV pulse.
    Finally we demonstrate that the bleaching of $\sivn$ induced by the DUV is accompanied by a dramatic increase in the neutral $\sivz$ population; $\sivz$ remains robust to extended periods of near-infrared excitation despite being a non-equilibrium state. DUV excitation thus presents a reliable method of generating $\sivz$, a desirable charge state for quantum network applications that is challenging to obtain by equilibrium Fermi engineering alone.
    Our results on two separate color centers at technologically relevant temperatures indicate a potential for above-band-gap excitation as a universal means of generating the neutral charge states of quantum point defects on demand.
\end{abstract}

\maketitle

\section{Introduction}

Optically active, quantum point defects in solid-state host crystals have shown promise as a platform to implement a wide range of quantum technologies~\cite{aharonovich2016sss, atature2018mps, bassett2019qdd, wolfowicz2021qgs}.
In most applications, it is ideal for the defect to remain in a particular charge state; e.g. the negatively charged nitrogen vacancy NV$^-$ for quantum sensing~\cite{schirhagl2014nvc} or the neutrally charged silicon vacancy $\sivz$ for quantum networking~\cite{rose2018oei}. 
Different charge states of the same defect have vastly different optical and spin properties which prevents the defect's use when in the incorrect state.

Control of defect charge states can be achieved through a diverse range of techniques involving chemical, electrical, and optical processes. 
Chemical techniques, such as passive Fermi-level engineering via bulk doping of the host lattice~\cite{radishev2021inv,doi2016pnc,grootberning2014pcs,rose2018oei} or functionalization of the surface for near-surface defects~\cite{hauf2011ccc,fu2010cnn,stacey2019eps,pederson2023otd,rodgers2024dsf} are limited by the challenge of obtaining both n-type and p-type doping in a single wide band gap material~\cite{walukiewicz1989and}, and rely on equilibrium conditions which may not be reached on practical timescales. Active electrical Fermi-level engineering~\cite{schreyvogel2015acs} similarly relies on equilibrium conditions. 
Even in cases in which Fermi-level engineering can be achieved, unintended ionization of the defect is often unavoidable so an active method of modifying the defect charge state \textit{in situ} is still necessary given the long non-equilibrium time constants. On the flip side, the long-lived nature of non-equilibrium states, opens up alternative methods for charge state control.  Non-equilibrium control can be achieved by optical, electrical~\cite{kato2013tle}, or combined optical/electrical~\cite{rieger2024foc} carrier generation. 
Intra-gap optical excitation has been used to demonstrate long-lived optical storage with the NV center~\cite{dhomkar2016ltd,monge2024rod} and stabilized SiV$^-$ emission after prolonged exposure to blue light~\cite{volker2024csd}. 
Furthermore, optical ionization requires no alteration of the sample, can be extremely rapid, and even tuned to selectively ionize a specific defect. 

Recently, selective intra-gap optical excitation has been used to ionize an ancillary defect to control the target defect charge state via photo-generated carriers~\cite{lozovoi2021oad,lozovoi2022idc,wood2023rtp,garciaarellano2024pic}. 
This technique, known as ``photodoping'', has revealed a large disparity between charged and neutral defect carrier capture cross-sections due to the Coloumbic attraction.
Negative NV$^-$ centers possess extremely large hole capture cross sections while neutral NV$^0$ show negligible charge capture~\cite{lozovoi2021oad}. 
These results suggest an interesting alternative charge state control method: the generation of free charge carriers by above-band-gap excitation which then preferentially neutralize all charged defects.

Utilizing above-band-gap excitation poses unique experimental challenges in diamond; diamond is an indirect semiconductor with a large band gap of 5.5\,eV corresponding to a deep-ultraviolet (DUV) wavelength of 225\,nm. 
Not only are excitation sources at these DUV wavelengths limited, but microscope optics optimized for visible or infrared (IR) are often absorptive at DUV. 
Furthermore, the dynamics of photo-injected carriers in diamond are complex, as tightly bound excitons form, diffuse, disassociate, and reform, generating free charge carriers which may be captured by defects~\cite{naka2016eft}. Nevertheless, a body of existing work has consistently demonstrated the effectiveness of DUV excitation at neutralizing defects in diamond by minutes-to-hours-long exposure to UV before measurement. Neutralization of N and NVH was observed after exposure to a Xe arc lamp~\cite{khan2009cte} and 215~nm laser~\cite{khan2013ccd}. 
DUV exposure was integral to assigning the 1.31\,eV emission to the neutral SiV center~\cite{johansson2011opn}.
These results suggest that DUV excitation is defect agnostic, unlike intra-band-gap excitation, and thus a unique tool for charge control in diamond. Still, many questions remain regarding the time-scale of ionization, power requirements, achievable charge state polarization, and ability to operate at technologically relevant temperatures, all of which are critical in establishing DUV excitation as a practical technique for quantum technologies.

In this work we demonstrate the integration of DUV excitation into a cryogenic visible/IR confocal microscope. We use this setup to demonstrate the effectiveness of DUV excitation in neutralizing the two most studied defects in diamond, NV and SiV, whose neutral and negative charge states can be directly detected through photoluminescence (PL). 
Time-resolved measurements reveal that the neutralization occurs on time scales competitive with conventional intra-gap photodoping, without the need for an ancillary defect. 
Furthermore, we utilize the technique as a means of studying the spectra of $\sivz$, a defect which exhibits long spin coherence times and stable, spin-dependent optical emission~\cite{rose2018oei,zhang2020odm}.
$\sivz$ has been historically difficult to observe due to its unusual temperature dependence, and tendency to bleach to $\sivn$ under direct intra-band-gap excitation. 

\section{Methods}

\begin{figure}
    \centering
    \includegraphics[]{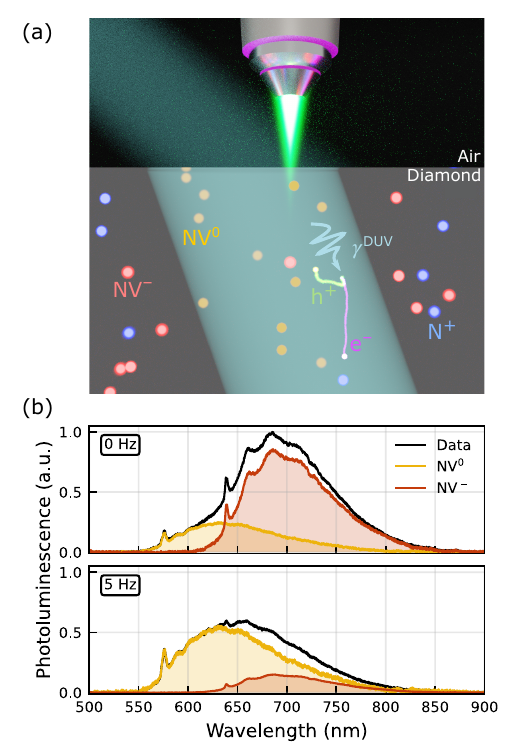}
    \caption{
        \textbf{Neutralization by DUV.}
        (a) An illustration of the experimental set up. 
        The DUV laser illuminates the sample from an oblique angle while a confocal microscope simultaneously performs a PL measurement.
        The DUV photon generates an electron and a hole in the diamond host lattice which are preferentially captured by charged defects (here NV centers and substitutional nitrogen) as assisted by the Coulombic interaction.
        (b) In Sample A, when the DUV laser is off (top) the signal is dominated by $\nvn$.
        When the DUV laser is on (bottom), the PL spectrum is dominated by $\nvz$.
        Spectra were taken under 8-\textmu W of 532-nm probe excitation at room temperature.
        Details of the spectral decomposition are discussed in SI.\SIrtnv.
    }
    \label{fig:methods}
\end{figure}

\begin{figure*}
    \centering
    \includegraphics{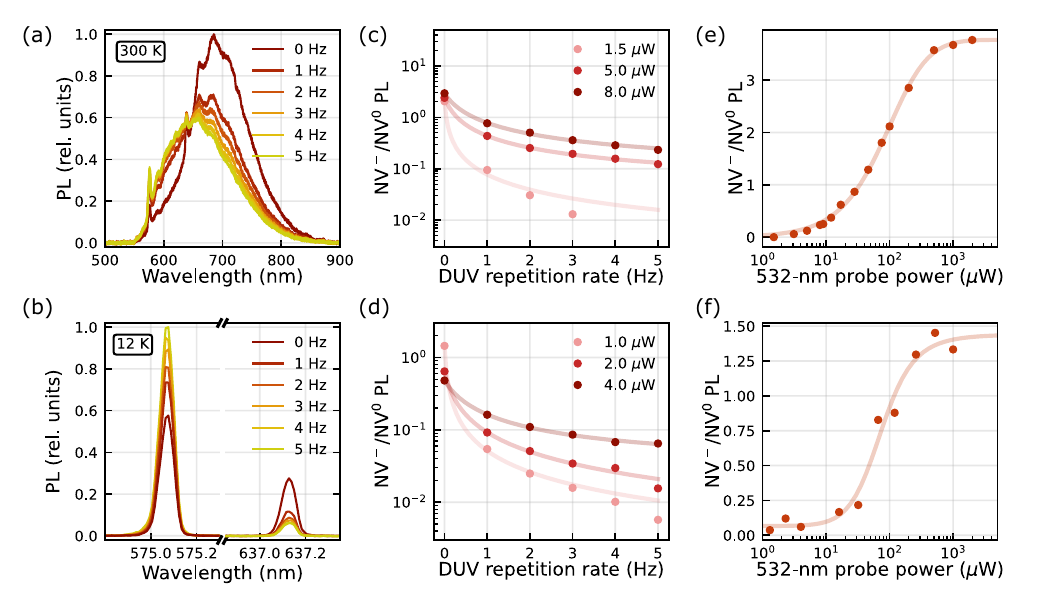}
    \caption{
        \textbf{Neutralization of the NV center.} 
        (a,b) NV center PL under 8-\textmu W 532-nm probe at different DUV pump repetition rates at room temperature and 12\,K respectively.
        A repetition rate of 0 Hz corresponds to the DUV laser being turned off.
        (c,d) Average PL intensity ratio of $\nvn/\nvz$ as a function of DUV repetition rate for different powers, also at room temperature and 12\,K respectively. 
        (e,f) Average PL intensity ratio of $\nvn/\nvz$ as a function of 532-nm probe power under 5-Hz DUV pump, also at room temperature and 12\,K respectively.
        In (c)--(f), the solid lines represent fits to the model Eq.~\eqref{eq:nv_population_ratio} with additional discussion in SI.\SImodel.
    }
    \label{fig:nv_neutralization}
\end{figure*}

The basic experimental setup is illustrated in Fig.~\ref{fig:methods}a.
A DUV HeAg pulsed laser (Photon Systems) emitting approximately 3\,\textmu J of 224.8\,nm light per 100-\textmu s pulse is focused onto the diamond sample at an oblique angle between the PL collection objective and sample surface to circumvent the high UV absorption coefficient of the microscope optics. When focused onto the diamond, the DUV forms an approximately (1\,mm$\times$2\,mm) elliptical spot. Off-resonant continuous-wave (CW) excitation (532\,nm for NV$^{0/-}$ and SiV$^-$; 830\,nm for SiV$^0$; spot sizes of approximately 0.7 and 1\,\textmu m respectively) and collection of PL are performed perpendicular to the sample surface through a visible-to-near-infrared (NIR) confocal microscope (SI.\SIsetup).
The DUV laser operates at discrete pulse rates tunable from 1--5\,Hz.

Two different single-crystal diamond samples grown via chemical vapor deposition (CVD) are studied. 
The PL spectra in sample A (SC Plate CVD, Element Six, [N] $<$ 1 ppm, [B] $<$ 0.05 ppm) is dominated by the NV center with no observable signal from SiV. Sample B is intentionally doped with 0.1--0.3\,ppm of silicon. It's spectra is dominated by SiV with no observable NV signal.

DUV photons incident on the diamond generate charge carriers by exciting valence electrons into the conduction band. 
During the subsequent, complex charge kinetics, some of the generated excitons and free charge carriers are trapped by impurities~\cite{takiyama1996pdk, naka2016eft}. 
Capture of charge carriers at deep impurities has been experimentally observed to significantly reduce the lifetime and diffusion length in doped diamond compared to higher purity samples~\cite{naka2016eft}. 
Due to the Coloumb interaction, electrons and holes are attracted to and are captured preferentially at positively and negatively charged defects respectively, generating neutral defects. 
The neutralization of charged defects by optical carrier generation has been experimentally reported in smaller band gap materials (e.g.~silicon) at low temperature, specifically to neutralize shallow donor qubits~\cite{saeedi2013rtq} in p-type material. 
While direct ionization by incident DUV photons is possible, the low power density of the DUV laser at the diamond sample suggests that direct ionization events are comparatively rare (SI.\SIbeamsize).

Fig.~\ref{fig:methods}b shows PL spectra of NV centers in Sample A, decomposed into the neutral $\nvz$ and negative $\nvn$ contributions, under DUV excitation.
Without the DUV (top panel), the signal is dominated by $\nvn$ with a small amount of $\nvz$ due to charge cycling induced by the 532-nm probe~\cite{bourgeois2017epd}.
With the DUV on (bottom panel), the signal is dominated by the $\nvz$ contribution, demonstrating the DUV neutralizing effect.

\section{NV center charge dynamics} \label{sec:nv_charge_dyanmics}
We first examine the quasi-steady-state neutralization of bulk NV centers in sample A under simultaneous exposure to the pulsed DUV pump and a 532-nm green probe.
The dominant charge states of the NV center, NV$^0$ and NV$^-$, both emit in the visible/NIR wavelength range (with zero-phonon line (ZPL) at 575\,nm and 637\,nm respectively) and can be optically excited into the phonon sideband (PSB) by 532\,nm excitation making the center an ideal probe for studying neutralization.

Figs.~\ref{fig:nv_neutralization}a,b show the quasi-steady-state spectra of the NV center under varying DUV laser repetition rates, at both 300\,K and 12\,K respectively.
In the absence of DUV pumping (0 Hz), the spectrum exhibits characteristic overlapping $\nvz$ and $\nvn$ contributions at roughly the same order of magnitude.
When the DUV is turned on at even the lowest repetition rate (1\,Hz), the spectrum exhibits a dramatic decrease in the $\nvn$ contribution, indicating significant neutralization of the defect.
Increasing the DUV repetition rate further results in increased $\nvz$ contribution, limited by the probe-induced charge cycling from $\nvz$ to $\nvn$. At the 5 Hz DUV rate, the probe-only time is 2000$\times$ longer than the DUV pulse time. 

To quantitatively study the effect of the DUV pulse, we perform a fitting procedure to determine the relative contributions of $\nvn$ and $\nvz$.
At low temperatures this is achieved by fitting of the respective ZPLs to a Voigt lineshape to remove the background and then integrating.
At room temperature, extracting the relative contributions of the two charge states is complicated by the overlap of the $\nvz$ and $\nvn$ signals.
While several techniques exist for decomposing composite spectra of this type~\cite{alsid2019pda}, we instead assume that the spectrum corresponding to the lowest probe power (1.5\,\textmu W) and highest DUV repetition rate (5\,Hz) constitutes a pure NV$^0$ signal.
This assumption is consistent with the lack of a detectable $\nvn$ ZPL at 637\,nm up to a relative contribution of $\nvn$ of less than 0.1\% (SI.\SIrtnv).
Utilizing this assumption, we are able to decompose the measured spectra into the corresponding $\nvn$ and $\nvz$ ``basis'' spectra (normalized to have equal integral, i.e. total counts).
Further details are provided in SI.\SIrtnv. 
The ratio of $\nvn$ to $\nvz$ PL intensities, $I_{\mathrm{NV^-}}/I_{\mathrm{NV^0}}$, under quasi-steady-state illumination of the CW 532-nm laser and pulsed DUV are shown in Fig.~\ref{fig:nv_neutralization}c--f.
The PL intensity ratio is proportional but not equal to the population ratio $\ev{N_{\nvn}}/\ev{N_{\nvz}}$ due to the difference in emission properties of both charge states.
We assume $I_{\mathrm{NV^-}}/I_{\mathrm{NV^0}}\approx 2.5\ev{N_{\nvn}}/\ev{N_{\nvz}}$~\cite{alsid2019pda} which we verify in SI.\SIrtnv.

Figs.~\ref{fig:nv_neutralization}c,d show the dependence on the DUV laser repetition rate for several 532-nm probe powers at 300\,K and 12\,K respectively.
Figs.~\ref{fig:nv_neutralization}e,f similarly show the dependence on the probe power at a constant (5 Hz) DUV repetition rate. We observe the same qualitative behavior at both temperatures; the NV signal becomes increasingly neutralized with higher DUV repetition rates and lower probe powers. 

We model the population dynamics under DUV and CW probe excitation using an empirical rate equation model (SI.\SImodel).
We find that the population ratio can be described by
\begin{equation} \label{eq:nv_population_ratio}
    \frac{\ev{N_{\nvn}}}{\ev{N_{\nvz}}} =
    \frac{
        \Gamma^{\mathrm{DUV}}_-\delta + \gamma^{\mathrm{eff}}_- T 
    }{
        \Gamma^{\mathrm{DUV}}_+\delta + \gamma^{\mathrm{eff}}_+ T 
    },
\end{equation}
where $\Gamma^{\mathrm{DUV}}_\pm$ and $\gamma^{\mathrm{eff}}_\pm$ are the DUV- and probe-induced charge conversion rates, whereas $\delta$ and $T$ are the DUV pulse length and pulse period respectively.
The subscript $\pm$ indicates if the corresponding process increases ($\nvn \to \nvz$) or decreases ($\nvz \to \nvn$) the charge of the NV center.
These rates encompass not only direction ionization (e.g. $\nvn \to \nvz + e^-$) but also charge capture (e.g. $\nvz + e^- \to \nvn$).
This model accurately describes the general dependence of the population ratio as a function of the DUV pulse rate and green power as shown in Figs.~\ref{fig:nv_neutralization}c--f.
Details of the model and its fitting to the data are discussed in depth in SI.\SImodel.

While the model can accurately describe the qualitative features of the population dynamics, it is phenomenological and thus incapable of distinguishing between specific processes.
Nevertheless we are able to determine the relative rates in aggregate.
Specifically we find that the DUV-mediated charge neutralization (which converts $\nvn \to \nvz$) occurs at a rate of up to five orders of magnitude faster than either of the probe-mediated conversion processes (which are generally within one order of magnitude of each other; SI.\SImodel).
This is despite the at least 100-fold lower DUV excitation intensity.
Additionally we find that the population dependence on the probe power (Fig.~\ref{fig:nv_neutralization}e,f) is largely dominated by one-photon-like processes (which scale linearly with probe power), indicating that the probe-mediated charge conversion occurs primarily through the capture of charges emitted from other defects (e.g. substitutional nitrogen).
These results indicate that the overall effect of the DUV can be straightforwardly understood despite the complexity of the NV center's charge dynamics and its interactions with other defects.

\begin{figure}
    \centering
    \includegraphics{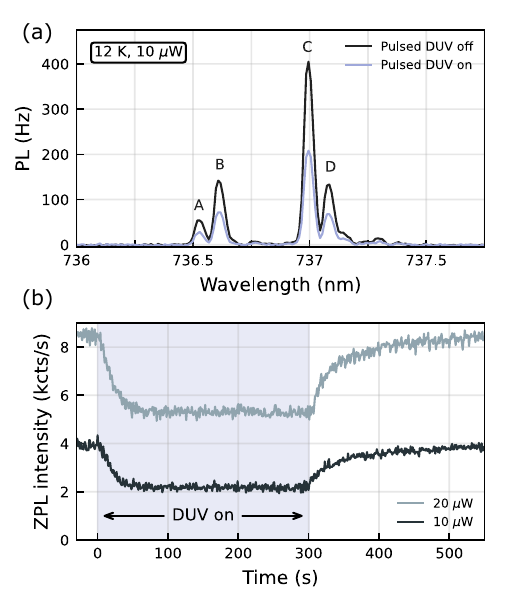}
    \caption{
        \textbf{Bleaching of $\sivn$ PL.}
        (a) Low-temperature (12\,K) PL of $\sivn$ centers under 10-\textmu W 532-nm probe with and without the DUV pump (1 Hz).
        For this power, the SiV PL is reduced by approximately a factor of 2 under DUV pumping evenly across the four primary transitions labeled A--D.
        (b) Integrated ZPL intensity as a function of time, initially starting with the DUV pump off ($t<0$\,s).
        The DUV pump (1\,Hz) is turned on at approximately $t=0$ causing a slow bleaching of the ZPL until a new equilibrium is established after about 30\,s, or approximately 30 pulses.
        At approximately $t=300$\,s, the DUV pump is turned off and a slow exponential-like recovery of the $\sivn$ PL is observed.
    }
    \label{fig:sivn_bleaching}
\end{figure}

\begin{figure}
    \centering
    \includegraphics{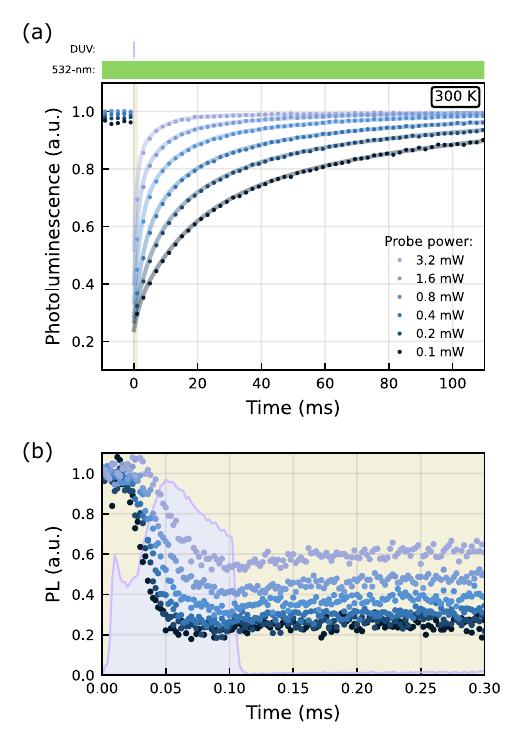}
    \caption{
        \textbf{Time dynamics of $\sivn$ recharging.} 
        (a) Time-resolved photoluminesence intensity of $\sivn$ under DUV pulsing at 5\,Hz and CW 532-nm probe excitation at various powers, corresponding fit to a triple-exponential (solid lines).
        The corresponding time constants span two orders of magnitude from 1--100\,ms.
        The plotted points (circles) are sparse samples of a lower-resolution binning (2-ms resolution) to aid in readability.
        Additional information about the fit is included in SI.\SIsivn.
        (b) Bleaching of the PL signal induced by the DUV pulse near time $t=0$\,s (indicated in (a) by the gold-colored stripe).
        The DUV pulse, shown in purple, lasts for 100\,\textmu s and was measured by excitation of $\nvz$.
    }
    \label{fig:sivn_recovery}
\end{figure}

\section{Time-resolved dynamics of SiV$^-$}\label{sec:sivn_bleaching}

\begin{figure*}
    \centering
    \includegraphics{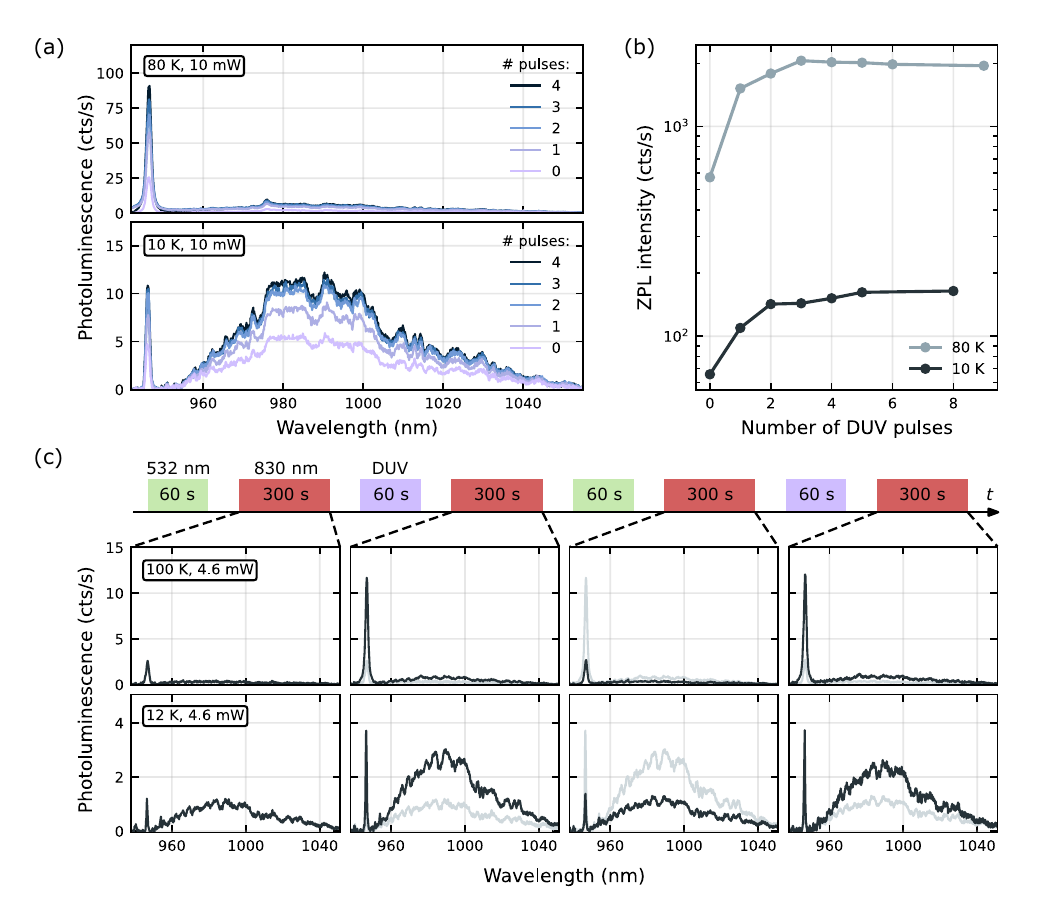}
    \caption{
        \textbf{Neutralization of SiV.} 
        (a) A series of $\sivz$ spectra taken with a 10-mW 830-nm probe laser after sequental exposure to single DUV pulses at 10\,K.
        (b) Fitted peak intensity of the ZPL line at 945-nm as function of the number of DUV pulses.
        We observe similar qualitative dependence at both 10\,K and 80\,K.
        (c) Sequence of spectra after alternating exposure to 532-nm and DUV.
        The $\sivz$ PL cycles between dark (after 532-nm exposure) to bright (after DUV exposure) repeatedly.
    }
    \label{fig:siv0}
\end{figure*}

We now turn our attention to Sample B which contains bulk SiV centers.
Compared to the NV center, these measurements are complicated by the inability to excite both charge states simultaneously with the same laser source~\cite{johansson2011opn,zhang2020odm}, and additional participating charge states~\cite{wood2023rtp, breeze2020dcs}.
We thus examine $\sivn$ and $\sivz$ separately, first considering the negatively charged $\sivn$.

Fig.~\ref{fig:sivn_bleaching}a shows the quasi-steady-state PL of the $\sivn$ ZPL lines under a CW 532-nm probe (10\,\textmu W) and DUV pulsed laser (5 Hz) at 12\,K.
We observe the suppression of the $\sivn$ ZPL intensity by about a factor of 2, indicating bleaching of the $\sivn$ into an alternative charge state.
While the specific charge state being generated cannot be ascertained from this measurement alone, subsequent measurements (Sec.~\ref{sec:sivz}) strongly suggest the bleaching corresponds to neutralization into $\sivz$.

Fig.~\ref{fig:sivn_bleaching}b shows a time trace of the $\sivn$ ZPL intensity as the DUV is turned on (at $t=0$\,s) and off (at $t=300$\,s).
Immediately after starting the DUV, the ZPL drops in intensity, reaching a new quasi-equilibrium after about 30\,s, wherein the intensity has dropped by about a factor of two.
Turning off the DUV at around $t=300$\,s initiates an exponential-like recovery of the ZPL which is understood to be a result of probe-induced re-ionization into $\sivn$.
Similar behavior is observed in the empirical rate equation model used to describe the NV center measurements (SI.\SImodel).

We investigate the recovery behavior of $\sivn$ using time-correlated PL measurements at room temperature.
The DUV laser, set to 5-Hz repetition rate, triggers the start of a detection window over which the $\sivn$ PL is monitored.
Fig.~\ref{fig:sivn_recovery}a shows the time-resolved PL signal for the first 100\,ms after the DUV pulse (at time $t=0$\,s).
After a rapid initial decay of the $\sivn$ PL during the 100-\textmu s DUV pulse, we observe an exponential-like recovery of the PL intensity with a strong dependence on the probe laser power.
We find that a triple-exponential curve accurately describes the PL recovery while also minimizing the number of parameters.
The fitted time constants span the full measurable range of 1--100\,ms which suggests that the recovery is complex, likely including multiple processes.
Similarly non-exponential time evolution has been observed in photodoping with sub-band-gap excitation~\cite{garciaarellano2024pic}.

Fig.~\ref{fig:sivn_recovery}b shows the same time-resolved PL in higher resolution during and immediately after the DUV pulse.
We observe an up to 80\% reduction in the $\sivn$ PL over the duration of the 100\,\textmu s pulse (shown in purple).
This suggests that the DUV can rapidly modify the defect charge states; the comparatively long DUV exposure times used in prior work is not required~\cite{khan2009cte,khan2013ccd,johansson2011opn}.

\section{Neutralization of SiV}\label{sec:sivz}
We perform PL measurements of $\sivz$ to confirm $\sivn$ is neutralized by the DUV.
We monitor the $\sivz$ population via PL excited by an 830\,nm CW diode laser.
The 830-nm excitation is unable to directly ionize $\sivz$ as it falls below the 1.5\,eV photoionization threshold~\cite{johansson2011opn,zhang2020odm}; nor is it able to ionize carriers from other defects such as vacancies, nitrogen-vacancies, or substitutional nitrogen~\cite{bourgeois2017epd}.
This enables the use of prolonged 830-nm excitation, even at high powers exceeding 1\,mW (SI.\SIsivz).

We utilize the CW 532-nm laser as a pump to initialize the SiV into $\sivn$ by exposing the sample at 1\,mW for 60\,s before reading out the $\sivz$ population with the 830-nm laser (10\,mW).
We next drive the sample with a single DUV pulse before measuring out the population again, repeating this process until the signal asymptotes to its maximal value.
The corresponding spectra are plotted in Fig.~\ref{fig:siv0}a showing a significant enhancement of the $\sivz$ ZPL peak by over a factor of 2 after only two DUV pulses. 
Fig.~\ref{fig:siv0}b shows the ZPL intensities as a function of the number of DUV pulses at both 10\,K and 80\,K which have very similar trends.
The rapid saturation of the ZPL intensity after the first few pulses is consistent with the rapid bleaching of the $\sivn$ signal shown in Fig.~\ref{fig:sivn_recovery}.
Details of the $\sivz$ spectral analysis are discussed in SI.\SIsivz.

The $\sivz$ spectra are markedly different at 10 and 80\,K.
Specifically, the 946-nm ZPL is greatly enhanced and the phonon sideband seemingly diminished at elevated temperatures.
Such behavior is explained by the presence of two excited states, $^3$E$_u$ and $^3$A$_{2u}$, corresponding to the 946-nm ZPL and 951-nm ZPL respectively, that have been directly observed in uni-axial stress measurements~\cite{green2019esn}. 
Due to the 6.8\,meV splitting, the $^3$E$_{u}$ state becomes frozen out at low temperatures, and thus its 946-nm ZPL and relatively small phonon sideband drop in intensity.
The large phonon sideband seen at 10\,K comes instead from the 951-nm ZPL emission from the $^3$A$_{2u}$ state which has a higher population at this temperature.

We next examine the repeatability of the photo-induced charge cycling.
Fig.~\ref{fig:siv0}c shows the $\sivz$ spectra after repeated cycling between 300-s PL measurements with the 830-nm probe (4.6\,mW) and 60-s pump steps using either 532-nm or DUV pumps to ionize or neutralize the SiV respectively.
We observe nearly identical spectra after a full cycle indicating that the population can be reliably initialized into $\sivz$ by the DUV excitation.
Such repeatable initialization, in combination with the stability of $\sivz$ under 830-nm excitation (SI.\SIsivz), demonstrates the viability of DUV exposure as a fundamental tool for $\sivz$-based quantum technologies.

\section{Conclusion and outlook}
We have demonstrated fast and on-demand neutralization of NV and SiV centers in diamond via above-band-gap, DUV excitation.
Using the DUV to reliably neutralize the defects, we then investigated the charge-state dynamics of the NV and SiV centers under combinations of CW probe excitation and DUV pumping.
These results, and their agreement with our modeling, indicate that the effect of the DUV is well understood and useful for defect-based applications. 
The technique also directly generates the neutral $\sivz$ without high-concentration co-doping which may accelerate the development of $\sivz$-based quantum nodes.

Beyond NV or SiV, this technique is agnostic to the structure and type of defect and may consequently be applied to study the neutral charge states of other interesting color centers such as the group-IV germanium-, tin-, and lead-vacancy centers.
This effect is also not limited to optically active defects but might also be used to similarly neutralize nearby, unintentionally-doped defects present in the host lattice which could contribute to electric-field noise. 
This can even be applied to neutralize the environment of charged color centers provided low-energy intra-gap excitation to re-initialize the target charged center minimally perturbs the neutralized environment. 
The neutralized environment would have reduced electric-field noise which has been associated with spectral diffusion and inhomogeneous broadening~\cite{tamarat2006ssc}.
Finally, we remark that this technique will be applicable in other emerging quantum defect host lattices~\cite{wolfowicz2021qgs, bassett2019qdd}.

\section*{Acknowledgments}
The SiV work was primarily supported by Department of Energy, Office of Science, National Quantum Information Science Research Centers, Co-design Center for Quantum Advantage (C2QA) under contract number DE-SC0012704.
The NV work was primarily supported by the National Science Foundation through the University of Washington Materials Research Science and Engineering Center, DMR-2308979.
N.S.Y.\ was supported by the National Science Foundation Graduate Research Fellowship Program under Grant No.~DGE-2140004. 
We thank Nathalie de Leon, Carlos Meriles and Chris Van de Walle for helpful discussions.

\bibliography{main}
\end{document}


\title{Supplementary Information for ``Rapid, in-situ neutralization of nitrogen- and silicon-vacancy centers in diamond using above-band-gap optical excitation''}

\author{Christian Pederson}%
\thanks{These two authors contributed equally}
\email{cpederso@uw.edu}
\affiliation{University of Washington, Physics Department, Seattle, WA, 98105, USA}%

\author{Nicholas S. Yama}%
\thanks{These two authors contributed equally}
\email{nsyama@uw.edu}
\affiliation{University of Washington, Electrical and Computer Engineering Department, Seattle, WA, 98105, USA}%

\author{Lane Beale}%
\affiliation{University of Washington, Physics Department, Seattle, WA, 98105, USA}%

\author{Matthew L. Markham}%
\affiliation{Element Six, Global Innovation Centre, Fermi Avenue, Harwell Oxford, Didcot, Oxfordshire, OX11 0QR, UK}%

\author{Kai-Mei C. Fu}%
\affiliation{University of Washington, Electrical and Computer Engineering Department, Seattle, WA, 98105, USA}%
\affiliation{University of Washington, Physics Department, Seattle, WA, 98105, USA}%
\affiliation{Physical Sciences Division, Pacific Northwest National Laboratory, Richland, Washington 99352, USA}

\maketitle            

\tableofcontents

\clearpage
\section{Experimental setup}

\begin{figure}[h]
    \centering
    \includegraphics[width=0.75\textwidth]{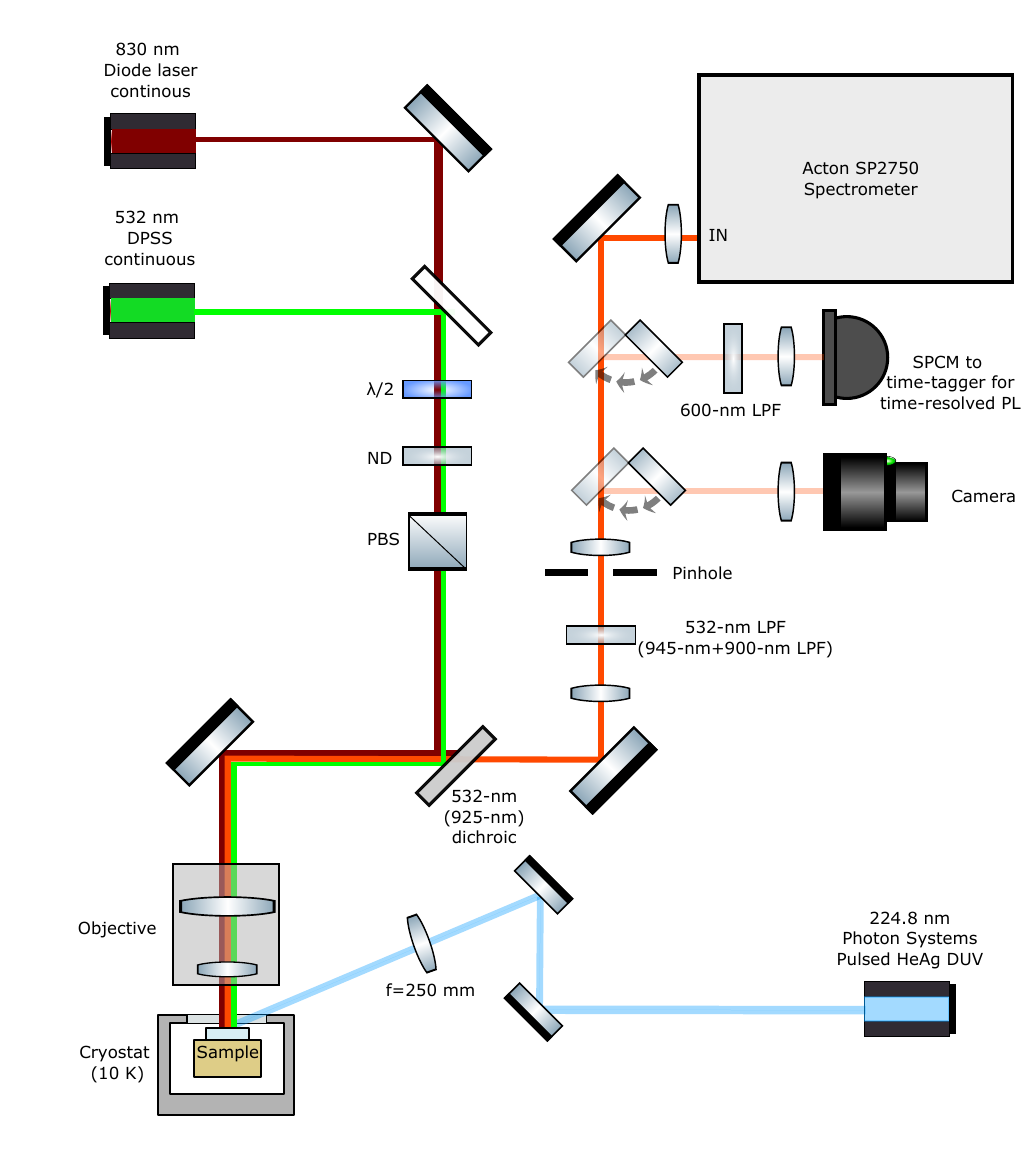}
    \caption{
    \textbf{Confocal microscope setup.}
    The samples are placed inside a closed-cycle He cryostat with quartz window (not shown).
    The DUV laser is focused via a 250-mm lens onto the sample at an oblique angle which bypasses the objective.
    A 532-nm(925-nm) dichroic mirror is used for the NV and $\sivn$($\sivz$) measurements respectively.
    The output path also uses either 532-nm long-pass filter (LPF) or combination of 900-nm and 925-nm LPFs depending on the defect and excitation.
    }
    \label{fig:system-schematic}
\end{figure}

\clearpage
\section{Estimation of DUV-induced photo-ionization rates and carrier densities}

\subsection{DUV beam properties}
The deep-ultraviolet (DUV) HeAg pulsed laser (Photon Systems) emits approximately 3\,\textmu J of 224.8\,nm light per 100-\textmu s pulse.
When aimed at a white index card, the laser excites photoluminescence (PL) in the visible range enabling an approximate measurement of the laser spot size and shape.
Near the aperture of the laser, the spot is imaged to be approximately 3\,mm in diameter and appears to have top-hat like profile.
A rough measurement of the spot at different lengths yields a relatively small divergence angle of $<1^\circ$ within the first meter after the aperture.

\subsection{Focused DUV spot size} 
\begin{figure}[h]
    \centering
    \includegraphics[width=0.75\linewidth]{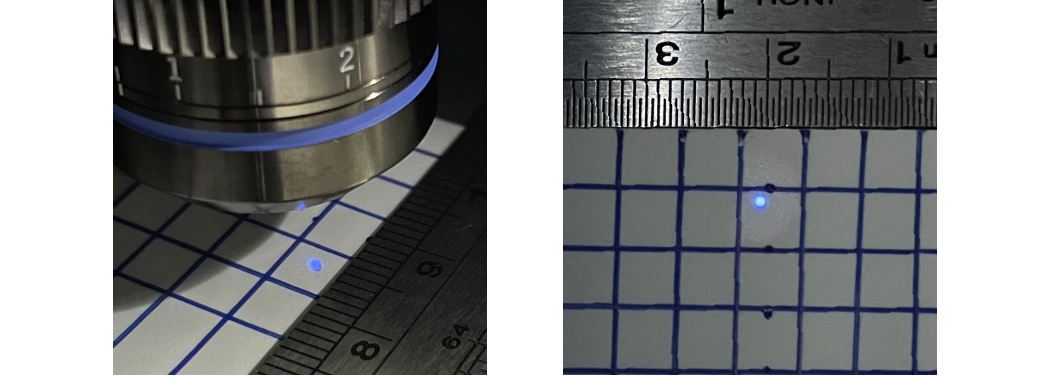}
    \caption{\textbf{DUV spot size.} 
    Images of the focused DUV spot on an index card on top of the cryostat (left) and when optimally focused at normal incidence (right).
    The blue lines are approximately 5\,mm apart and the ruler scale is in centimeters.
    }
    \label{fig:duv-spot-size}
\end{figure}

The focusing lens ($f = 250$\,mm) is placed roughly 50\,cm away from the aperture of the laser and 250\,mm from the top of the cryostat.
This focuses the DUV pulse into an elliptical spot on the top of the cryostat window with major and minor axis of approximately 2\,mm and 1\,mm in length respectively (Fig.~\ref{fig:duv-spot-size} left).
The oblique angle of the beam makes the system susceptible to errors in both the alignment and focusing of the lens, possibly increasing the spot size.
We determine an a lower limit on the focused spot size by replicating the DUV path without mirrors and at normal incidence to ensure proper alignment.
We find that optimally focused spot is approximately 0.9\,mm in diameter and has a depth of field of more than 10\,mm over which the spot size is roughly constant (Fig.~\ref{fig:duv-spot-size} right).

\subsection{DUV photon flux and power density at sample}
The DUV laser outputs approximately $E_{\mathrm{pulse}} \approx 3$\,\textmu J of energy per 100-\textmu s pulse.
A single DUV photon at 225\,nm has energy $E_\gamma = h c / \lambda = 9\times10^{-19}$\,J.
Thus the number of photons per pulse is $N = E_{\mathrm{pulse}} / E_\gamma \approx 3\times10^{12}$. 
Assuming a minimum circular spot spot of approximately 1\,mm in diameter at the sample, we obtain an upper bound on the photon flux of $I_{\mathrm{pulse}} =  0.04$\,photons/\r{A}$^2$.
However this does not account for losses due to reflection at the cryostat window and diamond surface.

Assuming an angle of incidence of approximately $50^\circ$ and a refractive index of $1.55$ (crystalline quartz), we calculate a reflection at the top surface (air-window interface) of the cyrostat window to be approximately 6.5\% for unpolarized light (increasing to as much as 12\% for $s$-polarized light).
Transmitted light refracts at an angle of approximately $29^\circ$ and travels through the roughly 500-\textmu m thick window with minimal absorption.
At the bottom surface (window-air interface), we find the reflection coefficient to also be 6.5\% for unpolarized light, which refracts back to $50^\circ$ degrees before reaching the sample.
We find that these values are generally insensitive to the angle of incidence within the range of 40--60$^\circ$.
Subsequent incidence at the diamond surface ($n = 2.717$ at 225\,nm) results in refraction at an angle of $16.4^\circ$ and reflection of about 22.5\%, which is similarly insensitive to the incidence angle.
Altogether, this corresponds to about 67\% of DUV photons transmitting into the diamond, or an upper limit on the photon flux of $I_{\mathrm{pulse}} \approx 0.03$\,photons/\r{A}$^2$.

\subsection{Direct ionization probability}
Assuming an ionization cross section of $\sigma = 10^{-1}$\,\r{A}$^2$ for the NV and SiV centers (extrapolated from \textit{ab initio} calculations~\cite{razinkovas2021pnc}), we may estimate the ionization probability per pulse as $P = \sigma I_{\mathrm{pulse}} \approx 3\times 10^{-3}$.
This small ionization probability cannot account for the observed charge conversion rates.

\subsection{Estimated exciton density}
\begin{figure}[h]
    \centering
    \includegraphics[]{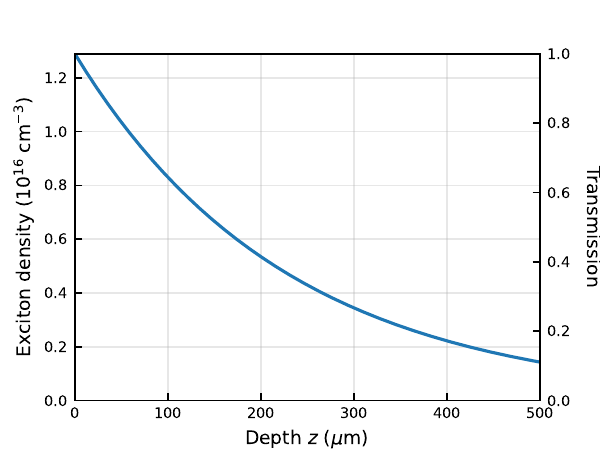}
    \caption{\textbf{Projected exciton density.}
    Calculated exciton density as determined by the photon density.
    }
    \label{fig:excition-density}
\end{figure}

The 224.8\,nm (5.515\,eV) excitation has sufficiently high energy to enable one-photon absorption within the diamond, generating an electron-hole pair.
The exciton density for the one-photon process is given by \cite{naka2008lte}
\begin{equation}
    n(z) = \alpha I_{\mathrm{pulse}} \exp(-\alpha z)
\end{equation}
where $\alpha = 44$\,cm$^{-1}$ \cite{clark1964iea} is the absorption coefficient, $I_{\mathrm{pulse}}=3\times 10^{14}$\,cm$^{-2}$ is the photon density per pulse, and $z$ is the depth into the sample.
This corresponds to an exciton density of $1.2\times10^{16}$\,cm$^{-3}$ near the surface of the sample as plotted in Fig.~\ref{fig:excition-density}.  
The absorption coefficient $\alpha$ is sample dependent, can vary significantly at the band edge, and has not been measured in our sample. 
Thus this calculation can only provide an estimate.
This exciton density is an upper bound for the free-carrier density and is comparable to defect densities (approximately $1.8\times10^{17}$\,cm$^{-3}$ nitrogen in Sample A and $3.6\times10^{16}$\,cm$^{-3}$ silicon for Sample B).

\clearpage
\section{Processing of room-temperature NV spectra}
\subsection{Pre-processing}
The room-temperature (RT) NV spectra were pre-processed to remove outliers (e.g. cosmic rays) by performing a rolling median over 30 pixels (approximately 3\,nm) and performing linear interpolation to replace points deviating by more than $\pm1.5\sigma$ of the 30-pixel sample ($\sigma$ is the standard deviation of rolling window samples).
This has an additional effect of reducing the noise and smoothing the data, however since this is performed on all spectra it should not impact the decomposition.
The uniform background due to by the CCD background offset, read noise, and dark counts is also removed at this stage.

\subsection{Determination of NV spectrum basis functions}
\begin{figure}[h]
    \centering
    \includegraphics[]{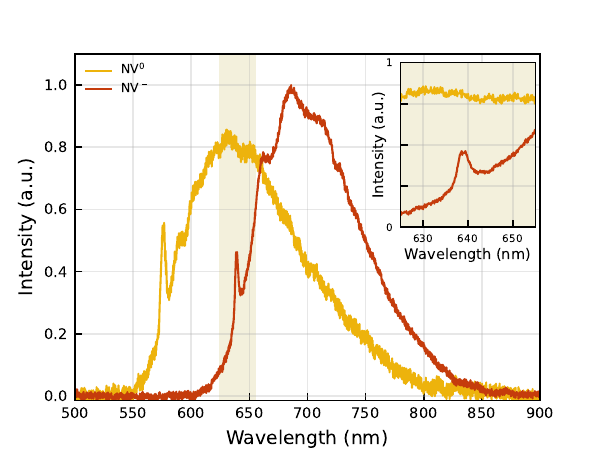}
    \caption{\textbf{NV basis spectra.}
        NV basis spectra normalized to unit intensity for fitting.
        The highlighted portion is shown in greater resolution in the inset, from which we do not observe any significant $\nvn$ ZPL in the $\nvz$ signal.
    }
    \label{fig:nv-basis}
\end{figure}

The RT NV spectrum (averaged over many DUV cycles) was qualitatively observed to contain significantly more $\nvz$ contribution at lower 532-nm probe powers and higher DUV repetition rates.
Although there exist sophisticated methods for decomposing NV center PL \cite{alsid2019pda}, the spectra with lowest probe power (1.5\,\textmu W) and highest DUV repetition rate (5\,Hz) is taken as the $\nvz$ basis spectrum $I_{\nvz}(\lambda)$ based off of the lack of a visible $\nvn$ ZPL peak near 637\,nm (Fig.~\ref{fig:nv-basis} inset).
The $\nvn$ basis spectrum was then obtained by summing all of the measured spectra $I_{\mathrm{tot}}(\lambda)$ and solving the minimization problem
\begin{equation}
    a_* = \underset{a}{\mathrm{argmin}} \int_{500}^{600}\dd{\lambda} |I_{\mathrm{tot}}(\lambda) - aI_{\nvz}(\lambda)|
\end{equation}
from which we define the $\nvn$ basis function as $I_{\nvn} = I_{\mathrm{tot}}(\lambda) - a_*I_{\nvz}(\lambda)$.
In other words, we determine the scaling of the $\nvz$ basis that minimizes the counts below 600\,nm (where $\nvn$ does not emit) and define the remaining signal as the $\nvn$ basis.
Both basis functions $I_{\nvz}(\lambda)$ and $I_{\nvn}(\lambda)$ are shown in Fig.~\ref{fig:nv-basis}.
To avoid issues of relative brightness between the two charge states we normalize each of the basis functions so that the counts are equal when integrated over the full range from 500-900\,nm.

\subsection{Decomposition of the room-temperature NV spectrum}
A generic NV spectrum $I_{\mathrm{NV}}(\lambda)$ is assumed to be composed of the $\nvn$ and $\nvz$ basis functions
\begin{equation}
    I_{\mathrm{NV}}(\lambda) = a I_{\nvz}(\lambda) + b I_{\nvn}(\lambda)
\end{equation}
where $a$, $b$ are constants.
We perform a fit of the measured signal to this model to determine the weighting parameters $a$ and $b$.
Overall we find good agreement between the data and the fit for all measured spectra.

\begin{figure}[t]
    \centering
    \includegraphics[]{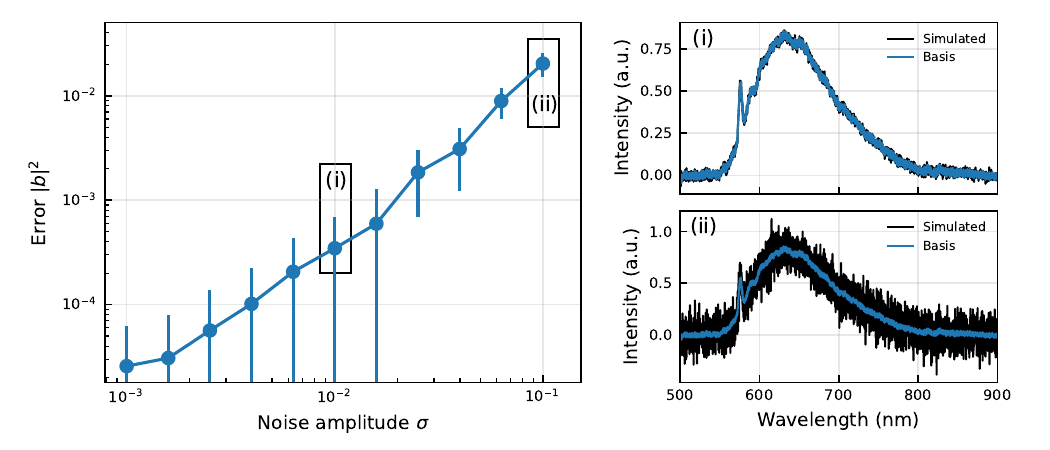}
    \caption{\textbf{Fitting error of NV$^0$ basis.}
        Estimated $\nvn$ component in the $\nvz$ basis function with added noise of standard deviation $\sigma$.
        (i) Plot of a representative simulated spectrum with $\sigma = 0.01$.
        It is of similar scale to the intrinsic basis function noise.
        (ii) Plot of representative simulated spectrum with $\sigma=0.1$ which is corresponds to relative noise markedly larger than any of the measured data.
    }
    \label{fig:fitting-error-1}
\end{figure}

The noise present in the basis functions can introduce uncertainty into the fitting procedure and so we aim to characterize its effect.
First we generate new $\nvz$ spectra by adding Gaussian noise $n(\lambda,\sigma)$ so that $I_{\nvz}^\prime (\lambda) = I_{\nvz}(\lambda) + n(\lambda)$, and then fit this new ``simulated'' spectra using the previously described process.
Repeating this for different noise amplitudes $\sigma$ (the standard deviation of the noise amplitudes) multiple times we can determine the fitting error which is defined as the erroneously fitted contribution of $\nvn$, i.e. $|b|$.
The error is plotted in Fig.~\ref{fig:fitting-error-1}.

\begin{figure}[]
    \centering
    \includegraphics[]{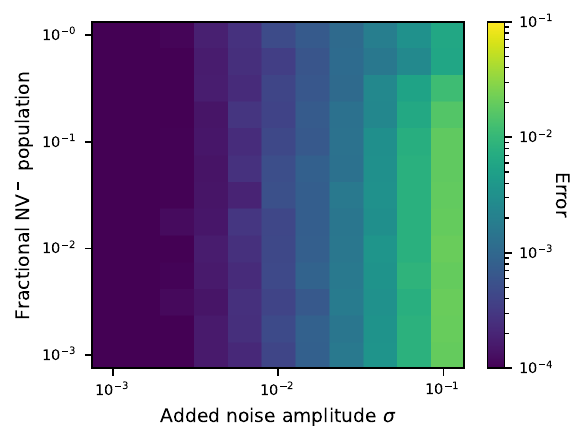}
    \caption{\textbf{Simulated fitting error of NV spectra.}
        Simulated error in estimation of $\nvn$ contribution as a function of the fractional $\nvn$ contribution and added noise $\sigma$.
        Error appears to be primarily limited by the added noise.
    }
    \label{fig:fitting-error-2}
\end{figure}

Because the basis function already contains noise with some standard deviation $\sigma^\prime$, when $\sigma \ll \sigma^\prime$ the new spectrum $I_{\nvz}^\prime (\lambda)$ is effectively identical and so the fit converges with increasing higher accuracy at smaller $\sigma$.
However, when the noises are of similar amplitude $\sigma \approx \sigma^\prime$ the simulated spectrum has the same signal-to-noise ratio (SNR) as the original and can be thought of as a repeated measurement.
We can use this ``effective'' repeated measurement to gauge the robustness of our basis functions.
We find that the noises are matched when $\sigma\approx10^{-2}$.
The corresponding fitting error is $4\times 10^{-5}$ indicating high confidence in the basis function orthogonality.

We can repeat this same process, now allowing for finite $\nvn$ contribution $b > 0$.
The resulting estimation errors of the fitted parameters are shown in Fig.~\ref{fig:fitting-error-2}.
For noise of similar SNR to the measurement ($\sigma\approx10^{-2}$) we find that the approximate error is on the order of $10^{-3}$ or less for all fractional $\nvn$ populations (ranging from $10^{-3}$ to $10^0$).
This indicates that the fit accurately identifies fractional $\nvn$ populations of less than $1\%$ in the worst case, with increasing accuracy at higher SNR.

\subsection{Estimate of the intrinsic PL intensity ratio}
The basis functions $I_{\nvn}(\lambda)$ and $I_{\nvz}(\lambda)$ are normalized to have equal integral $\int\dd{\lambda}I_{\nvn}(\lambda) = \int\dd{\lambda}I_{\nvz}(\lambda)$.
However, due to the difference in relative brightness of $\nvn$ and $\nvz$ (under the same excitation), the basis functions themselves correspond to different numbers of $\nvn$ or $\nvz$ defects.
As a consequence the actual population ratio $\ev{N_{\nvn}}/
\ev{N_{\nvz}}$ is only proportional to the PL intensity ratio $b/a$.
The scaling has been determined (for 532-nm excitation) as $b/a\approx 2.5\ev{N_{\nvn}}/\ev{N_{\nvz}}$~\cite{alsid2019pda} which indicates that the PL of a single $\nvn$ is about 2.5-times brighter than a single $\nvz$.

We examine the relative brightness of $\nvn$ and $\nvz$ by comparing spectra at different DUV repetition rates.
Assuming that all NV centers remain in either $\nvn$ or $\nvz$; any increase in the $\nvz$ PL intensity (due to a change in DUV repetition rate) should be accompanied by a proportional decrease in the $\nvn$ PL intensity. 
The proportionality constant is then the relative brightness of $\nvn$ and $\nvz$.
However, if there exists other relevant NV charge states, one would expect the proportionality factor vary significantly at different DUV repetition rates.

\begin{figure}[h]
    \centering
    \includegraphics[]{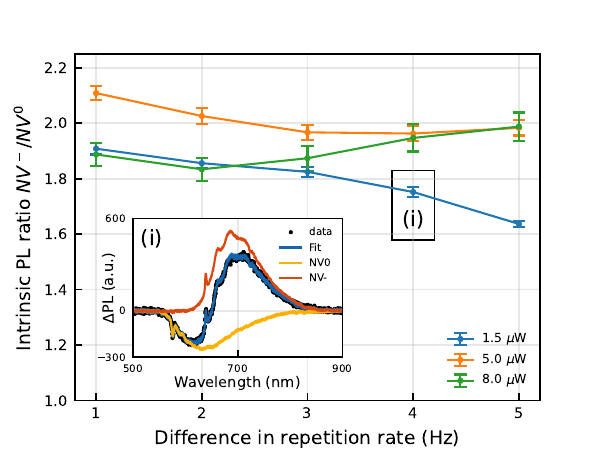}
    \caption{\textbf{Estimated intrinsic PL ratio.}
        Calculated intrinsic PL intensity $\nvn$/$\nvz$ intensity ratio versus difference in DUV repetition rates.
        The ratio is calculated by fitting the difference of the 0-Hz (DUV-off) spectrum and the spectrum corresponding to a given repetition rate as shown in the inset (corresponding to marked data point).
    }
    \label{fig:nv_prop_factor}
\end{figure}

Fig.~\ref{fig:nv_prop_factor} shows the proportionality constant obtained by fitting the difference of the 0-Hz PL signal (DUV off) and the PL at each DUV repetition rate for a given power.
Averaging all measured points we determine an intrinsic intensity ratio of approximately $1.8$ which roughly agrees with the literature value of $2.5\pm0.5$~\cite{alsid2019pda}.
The difference might be attributed to a wavelength-dependent difference in collection and detection efficiency which was not accounted for in this analysis.
The intrinsic intensity ratio appears to vary slightly with repetition rate, however no clear trend is observed between the different probe powers.

\clearpage
\section{Theoretical modeling}

\subsection{Rate equations}

The observed charge state dynamics could be modeled as number of different defect populations $N_j(\vb x, t)$ which interact via itinerant electrons $n(\vb x, t)$ and holes $p(\vb x, t)$ either by photo-ionization (e.g. $\nvn \to \nvz + e^-$) or charge capture (e.g. $\nvz + e^- \to \nvn$). 
In this model, the above-band-gap DUV excitation generates free charge carrier pairs ($\gamma^{\mathrm{DUV}} \to e^- + h^+$) which can then diffuse, eventually recombining or being captured by defects.
Provided all relevant interactions are known, one could construct a system of rate equations describing the full dynamics of the system.

In the NV system, the relevant populations can be assumed to be the two NV charge states, $\nvn$ and $\nvz$, and substitutional nitrogen, which also exists in two charge states $\Np$ and $\Nz$.
Taking into account charge cycling and capture processes, a system of equations can be written
\begin{subequations} \label{eq:nv_model_full}
    \begin{align} 
        \dot N_{\nvn} &=  -\gamma_- N_{\nvn} + \gamma_0 N_{\nvz} + K_0^e n N_{\nvz} - K_-^h p N_{\nvn},\\
        \dot N_{\nvz} &=  \gamma_- N_{\nvn} - \gamma_0 N_{\nvz} - K_0^e n N_{\nvz} + K_-^h p N_{\nvn},\\
        \dot N_{\Np} &=  \gamma_n N_{\Nz} - K_n^e n N_{\Np} + K_n^h p N_{\Nz}, \\
        \dot N_{\Nz} &=  -\gamma_n N_{\Nz} + K_n^e n N_{\Np} - K_n^h p N_{\Nz},\\
        \dot n &=  \gamma_- N_- - K_0^e n N_{\nvz} + \gamma_n N_{\Nz} - K_n^e n N_{\Np} + \Gamma_{\mathrm{DUV}}(t) - K_{eh} np, \\
        \dot p &=  \gamma_0 N_{\nvz} - K_-^h n N_{\nvn} - K_n^h p N_{\Nz} + \Gamma_{\mathrm{DUV}}(t) - K_{eh} np,  
    \end{align}
\end{subequations}
where the one-particle terms (e.g. $\gamma_-N_{\nvn}$) correspond to photo-ionization ($\nvn \to \nvz + e^-$) by the green probe and the two-particle terms (e.g. $K_0^e n N_{\nvz}$) correspond to absorption ($\nvz + e^- \to \nvn$).
In this model we neglect the spatial dependence due to our collection and excitation occurring at the same position.
Similar models have been used throughout the literature, often including additional interaction terms and populations \cite{wood2023rtp,garciaarellano2024pic}.
However, the success of the model is largely dependent on whether or not all relevant processes have been considered.
Not only are many of the relevant processes often unknown, but they are multitudinous --- often to the point of over-parameterization.

To avoid such issues, we make the assumption that the primary defect under consideration is sufficiently sparse (compared to other defects) such that its effect on the itinerant carrier populations is negligible.
This is particularly valid in the case of Sample A, in which substitutional nitrogen outnumbers NV by at least 100:1~\cite{edmonds2012pon}.
As a consequence, the electron and hole dynamics are primarily determined by the other (dominant) defect populations and DUV-induced carrier injection.
The resulting solutions $\tilde n(t)$ and $\tilde p(t)$ can then be inserted into the rate equations for the primary defect, in which they act as explicit time-dependent coefficients.
In the case of the NV center, this enables simplification of the full model Eqs.~\eqref{eq:nv_model_full} to
\begin{equation} \label{eq:nv_model_simplified}
    \dv{t}
    \begin{bmatrix}
        N_{\nvn} \\ N_{\nvz}
    \end{bmatrix}
    =
    \begin{bmatrix}
        -\Gamma_+(t) & \Gamma_-(t) \\
        \Gamma_+(t) & -\Gamma_-(t)
    \end{bmatrix}
    \begin{bmatrix}
        N_{\nvn} \\ N_{\nvz}
    \end{bmatrix},
\end{equation}
where $\Gamma_+(t) = \gamma_- + K^h_- \tilde p(t)$ and $\Gamma_-(t) = \gamma_0 + K^e_0 \tilde n(t)$ are the rates of raising ($\nvn \to \nvz$) and lowering ($\nvz \to \nvn$) the charge state respectively.
Finally, because the defect-specific processes have been abstracted out, this model could equally apply to other defects like the SiV, which we discuss in \ref{sec:siv_pl_dynamics_model}.

\subsection{Pulsed-pump-probe measurements} \label{sec:pulsed-pump-probe-measurements}
We now describe solutions of the model in the pulsed-pump-probe scheme. 
We take the rates be to piece-wise constant
\begin{equation}
    \Gamma_\pm(t) = 
    \begin{cases}
        \nu_\pm, & 0 \leq t < \delta \\
        \kappa_\pm, & \delta \leq t < T
    \end{cases},
\end{equation}
where $\nu_\pm$ and $\kappa_\pm$ are constants corresponding to the effective rates when the DUV is on and off respectively.
The DUV pulse length is given by $\delta$ and has period $T$.
A similar model for pulsed voltage measurements~\cite{pambukhchyan2023pns}.

We analytically solve Eq.~\eqref{eq:nv_model_simplified} within each domain $0< t < \delta$ and $\delta < t < T$, while enforcing continuity and periodicity.
Focusing on the first domain during the DUV pulse, $0 < t <\delta$, one finds that the eigenvalues of the coefficient matrix $\bm\Gamma$ are given by $\{0, -(\nu_+ + \nu_-)\}$ which indicate a steady state (corresponding to $N_{\nvn} / N_{\nvz} = \nu_-/\nu+$) and a transient decaying at a rate $(\nu_+ + \nu_-)$.
Solutions to the model Eq.~\eqref{eq:nv_model_simplified} within each domain are characterized by the time-evolution operator $\vb P_\nu(\Delta t) = \exp(\bm\Gamma \Delta t)$ which is written in the original basis as
\begin{equation}
    \vb P_\gamma (\Delta t) =
    \begin{bmatrix}
            \frac{\nu_-}{\nu_+ + \nu_-} + \frac{\nu_+}{\nu_+ + \nu_-}e^{-(\nu_++\nu_-)\Delta t} 
            & 
            \frac{\nu_-}{\nu_+ + \nu_-} - \frac{\nu_-}{\nu_+ + \nu_-}e^{-(\nu_++\nu_-)\Delta t}  
        \\
            \frac{\nu_+}{\nu_+ + \nu_-} - \frac{\nu_+}{\nu_+ + \nu_-}e^{-(\nu_+ +\nu_-)\Delta t} 
            & 
            \frac{\nu_+}{\nu_+ + \nu_-} + \frac{\nu_-}{\nu_+ + \nu_-}e^{-(\nu_++\nu_-)\Delta t} 
    \end{bmatrix}.
\end{equation}
An identical operator with $\nu_\pm \to \kappa_\pm$ can be obtained for the second domain while the DUV is off, completely characterizing the time evolution of the system.

\begin{figure}[h]
    \centering
    \includegraphics{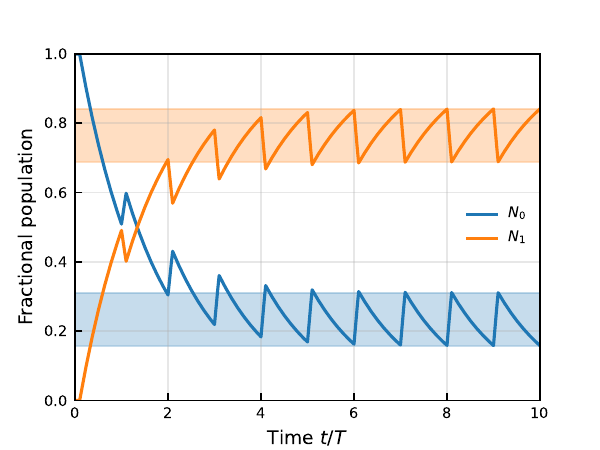}
    \caption{\textbf{Population dynamics of pulsed-pump-probe measurements.}
        Dynamics of the populations as a function of time (normalized to the period).
        The populations undergo interleaved periods of decay and growth corresponding to an initial DUV bleach (with respect to the orange curve) and slow probe recharging.
        The populations ultimately reach an equilibrium after a few cycles in which they oscillate between two extrema defined by the eigenvectors of the time-evolution operator Eq.~\eqref{eq:propagator_eigenvector}.
        This plot corresponds to parameters $\delta/T = 0.1$, $(T\nu_+,T\nu_-) = (0,2)$, and $(T\kappa_+,T\kappa_-) = (0.75,0)$.
    }
    \label{fig:pulsed-pump-probe-1}
\end{figure}

A representative example of the system population evolution is shown in Fig.~\ref{fig:pulsed-pump-probe-1}.
At fast time scales we observe characteristic exponential decay/growth of the populations, alternating between each domain as the DUV is turned off or on.
On slow time scales, the average population slowly grows or decays over multiple cycles, asymptotically reaching quasi-equilibrium when the effects of the DUV pulse and probe laser are balanced.
The quasi-equilibrium population at the start of the DUV pulse, $\vb N_*$, corresponds to the eigenvector of the full-period time-evolution operator $\mathcal P = P_\kappa(T-\delta) P_\nu(\delta)$ with eigenvalue unity.
One finds, up to suitable normalization, that
\begin{equation} \label{eq:propagator_eigenvector}
    \vb N_* = 
    \begin{bmatrix}
            -e^{(\kappa_+ + \kappa_-)T}(\nu_+ + \nu_-)\kappa_-
            + e^{(-\nu_+ - \nu_- + \kappa_+ + \kappa_-)\delta}(\kappa_+ + \kappa_-)\nu_- 
            + e^{(\kappa_+ + \kappa_-) \delta} (-\nu_-\kappa_+ + \nu_+\kappa_-)
         \\
            -e^{(\kappa_+ + \kappa_-)T}(\nu_+ + \nu_-)\kappa_+
            + e^{(-\nu_+ - \nu_- + \kappa_+ + \kappa_-)\delta}(\kappa_+ + \kappa_-)\nu_+ 
            + e^{(\kappa_+ + \kappa_-) \delta} (\nu_-\kappa_+ - \nu_+\kappa_-)
     \end{bmatrix}
\end{equation}
The state $\vb N_*$, along with the state after the DUV pulse $P_\nu(\delta) \vb N_*$, describe the extrema of the populations at quasi-equilibrium.

Experimentally we measure the average quasi-equilibrium populations of the $\nvn$ and $\nvz$ via PL.
Formally, this would be described by integrating $\vb N(t)$ over a full period assuming $\vb N(0) = \vb N_*$.
However, in the limit where the DUV pulse is short ($\delta \ll T$) and the green power is weak ($\kappa_j T \ll 1$), we can approximate the average population over the period as simply the average of the extrema, i.e. $[\ev{N_{\nvn}}, \ev{N_{\nvz}}] = (\vb N_* + \vb P_\nu(\delta) \vb N_*)/2$.
The quasi-equilibrium, average population ratio given then by
\begin{equation}
    \frac{\ev{N_{\nvn}}}{\ev{N_{\nvz}}} =
    \frac{
        (e^{\kappa T} - e^{(\kappa+\nu)\delta}) (\nu_- \kappa_+ - \nu_+ \kappa_-)
        +(e^{\kappa \delta} - e^{\nu\delta + \kappa T}) (\nu_+ \kappa_- + \nu_-(\kappa_+ + 2\kappa_-))
    }{
        -(e^{\kappa T} - e^{(\kappa+\nu)\delta}) (\nu_- \kappa_+ - \nu_+ \kappa_-)
        +(e^{\kappa \delta} - e^{\nu\delta + \kappa T}) (\nu_- \kappa_+ + \nu_+(2\kappa_+ + \kappa_-))
    }
\end{equation}
where $\nu = \nu_+ + \nu_-$ and $\kappa = \kappa_+ + \kappa_-$.

As a final step we make some further simplifications.
First we assume that the rates are sufficiently small so that the exponential factors can be linearized, $\exp(\kappa T) \approx 1 + \kappa T$ and $\exp(\nu \delta) \approx 1 + \nu\delta$.
Then, if the rates can be simplified as
\begin{align*}
    \nu_+ &= \gamma^{\mathrm{eff}}_+ + \Gamma^{\mathrm{DUV}}_+, & \nu_- &= \gamma^{\mathrm{eff}}_- + \Gamma^{\mathrm{DUV}}_-, \\
    \kappa_+ &= \gamma^{\mathrm{eff}}_+, & \kappa_- &= \gamma^{\mathrm{eff}}_- ,
\end{align*}
the ratio can be written 
\begin{equation} \label{eq:population_ratio_model}
    \frac{\ev{N_{\nvn}}}{\ev{N_{\nvz}}} \approx
    \frac{
        \Gamma^{\mathrm{DUV}}_-\delta + \gamma^{\mathrm{eff}}_- T 
    }{
        \Gamma^{\mathrm{DUV}}_+\delta + \gamma^{\mathrm{eff}}_+ T 
    }
\end{equation}
where $\gamma^{\mathrm{eff}}_\pm$ are effective rates of raising/lowering the charge due to the green probe (including both direct, photo-induced processes as well as those mediated by other defects) and $\Gamma^{\mathrm{DUV}}_\pm$ describe charge raising/lowering processes induced by the DUV.

\subsection{Application to NV center charge state dynamics}
As discussed in the main text, after normalizing the PL intensity ratio $I_{\mathrm{NV^-}}/I_{\mathrm{NV^0}}$ to account for the difference in intrinsic PL brightness, we then fit the data against the model Eq.~\eqref{eq:population_ratio_model}.
Notably, the model is over determined as all rates can be rescaled by a constant without changing the ratio.
We instead determine the relative rates normalized by $\gamma^{\mathrm{eff}}_-$ (corresponding to processes $\nvz \to \nvn$) which is expected to have contribution primarily from probe-induced hole ejection, as electron capture by $\nvz$ should be comparatively rare \cite{lozovoi2021oad}.
The fitting procedure is detailed below for sweeps of the repetition rate and probe power:
\begin{itemize}
    \item \textbf{Repetition rate sweep:} The repetition rate $r$ corresponds to the inverse of the period $T$ which serves as the independent variable of the fitting model.
    We directly fit to the function
    \begin{equation}
        f(x) = \frac{A + r^{-1}}{B + Cr^{-1}}
    \end{equation}
    where we can extract the relative rates as
    \begin{align*}
        \Gamma^{\mathrm{DUV}}_-/\gamma^{\mathrm{eff}}_- = A / \delta ,
        &&
        \Gamma^{\mathrm{DUV}}_+/\gamma^{\mathrm{eff}}_- = B / \delta ,
        &&
        \gamma^{\mathrm{eff}}_+/\gamma^{\mathrm{eff}}_- = C,
    \end{align*}
    where $r$ is of units Hertz and $\delta = 100$\,\textmu s.
    The resulting relative rates for the system at both 12 and 300\,K are shown in Fig.~\ref{fig:pulsed-pump-probe-2}.
    We observe that all rates appear to have similar relative values independent of the temperature.
    As expected, the DUV-induced hole capture $\Gamma^{\mathrm{DUV}}_+$ is more than three orders of magnitude larger than the probe-mediated processes for all measured powers.
    Curiously the DUV-mediated electron absorption terms (corresponding to $\nvz \to \nvn$) appear to rapidly turn on at an intermediate power of a few microwatts.
    While the abstraction of the model limits any definite statements, such behavior could be explained by the larger probe powers populating charge states of another defect species which aids in converting $\nvz \to \nvn$.

    \begin{figure}[h]
        \centering
        \includegraphics{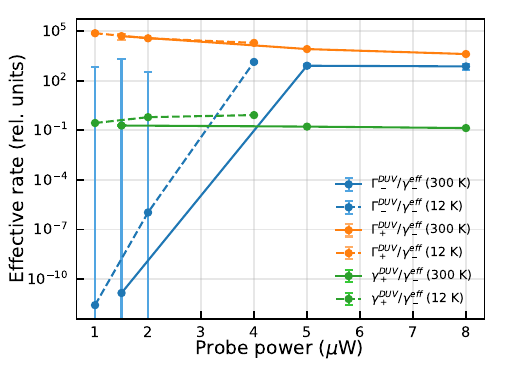}
        \caption{\textbf{Relative rates of the NV center charge dynamics.}
            Relative rates of the effective DUV- and green-related processes at both 12 and 300\,K, derived from the fit to the model.
            At low powers, the uncertainty of the DUV-induced $\nvz\to\nvn$ rate $\Gamma^{\mathrm{DUV}}_-$ is large, despite the fitted value being vanishingly small.
            This uncertainty persists because its effect on the population (given by $\delta\Gamma^{\mathrm{DUV}}_-$) is small but is masked by the short pulse length $\delta$.
        }
        \label{fig:pulsed-pump-probe-2}
    \end{figure}

    \item \textbf{Probe power sweep:} It is assumed that the direct ionization process between $\nvn$ and $\nvz$ (encapsulated by $\gamma^{\mathrm{eff}}_+$ and $\gamma^{\mathrm{eff}}_-$) is a two-photon process and depends quadratically on the power $p$.
    However, these rates also include the net effect of other defect species (such as substitutional nitrogen) which can be ionized via a one-photon process.
    Additionally, it is not possible to straightforwardly separate the power dependence of $\gamma^{\mathrm{eff}}_+$ and $\gamma^{\mathrm{eff}}_-$ from $\Gamma^{\mathrm{DUV}}_+$ and $\Gamma^{\mathrm{DUV}}_-$.
    As such we model the probe power sweep by the phenomenological model
    \begin{equation}
        f(P) = \frac{A + Bp + Cp^2}{1 + Dp + Ep^2}.
    \end{equation}
    The relative scale of the coefficients $\{A,...,E\}$ can then be used to assess the relative significance of one-/two-photon processes.
    As before, we normalize to one of the terms to obtain relative amplitudes.
    The resulting parameters for both measured temperatures are shown in Table~\ref{tab:nv_power_sweep}.

    \begin{table}[h]
        \centering
        \begin{tabular}{ c c c c c }
            \hline \hline
             & \qquad & 300\,K & \qquad & 12\,k \\
            \hline
            A (1) & \qquad\qquad\qquad
                & $0.000 \pm 0.079$  & \qquad\qquad\qquad
                & $0.17 \pm 0.23$ \\
            B (\textmu W$^{-1}$) &
                & $0.090 \pm 0.009$ &
                & $0.000 \pm 0.048$ \\
            C (\textmu W$^{-2}$)  &
                & $0.0003 \pm 0.0002$ & 
                & $0.001 \pm 0.002$ \\
            D (\textmu W$^{-1}$)  &
                & $0.009 \pm 0.001$ &
                & $0.006 \pm 0.020$ \\
            E (\textmu W$^{-2}$)  &
                & $0.00003 \pm 0.00002$ &
                & $0.0003 \pm 0.0006$ \\
            \hline \hline
        \end{tabular}
        \caption{\textbf{Power sweep fit parameters}}
        \label{tab:nv_power_sweep}
    \end{table}
    
    We determine that at room temperature ($T=300$\,K) and low excitation powers ($<100$\,\textmu W), the charge conversion is dominated by linear, one-photon-like processes.
    This suggests that ionization is mediated primarily through another defect at low powers.
    
    At low temperatures ($T=12$\,K) we instead find negligible linear component of the $\nvz\to\nvn$ terms while the linear component of the $\nvn\to\nvz$ process is of similar scale to the room-temperature fit.
    While this might suggest that the single-photon $\nvz\to\nvn$ process freezes out at low temperatures, it is also possible that the responsible defect species (which contributes to $\nvz\to\nvn$) was not present in the spot(s) probed in the low-temperature data.
    This might also explain the observed difference in the NV charge-state ratios (with the 12-K data having significantly more $\nvz$).
\end{itemize}

\subsection{Relation to SiV PL dynamics} \label{sec:siv_pl_dynamics_model}
\begin{figure}[h]
    \centering
    \includegraphics{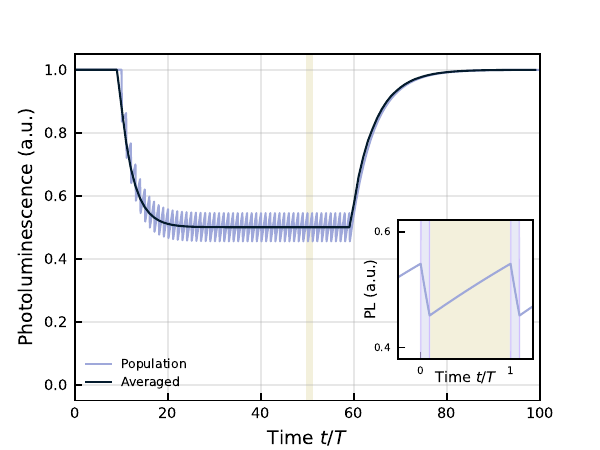}
    \caption{\textbf{Model of SiV$^-$ PL dynamics.}
        Simulated population dynamics and PL signal replicating the measured $\sivn$ PL dynamics of Figs.~3,4 in the main text.
        The DUV is ``turned on'' at $t/T=10$ and ``turned off'' at $t/T=60$.
        The population oscillates quickly, but the measured PL is the rolling average of the population over one (or more) periods $T$.
        The inset shows the population evolution over a single period in quasi-equilibrium with the simulated DUV pulses shown.
        This plot corresponds to parameters $\delta/T = 0.1$, $(T\nu_+,T\nu_-) = (1.8,0)$, and $(T\kappa_+,T\kappa_-) = (0, 0.2)$.
    }
    \label{fig:pulsed-pump-probe-3}
\end{figure}

Although the model was derived as a simple version of the NV charge state dynamics, it can also be used to qualitatively understand the behavior of the SiV system.
Fig.~\ref{fig:pulsed-pump-probe-3} shows the population of one of the levels as a function of time plotted with the rolling average over a single period (as is measured by PL).
This corresponds to the time-resolved $\sivn$ PL measurement shown in Fig.~3b.
At higher temporal resolution (inset), one finds contrasting periods of exponential decay and growth over the course of a particular cycle roughly corresponding to Fig.~4a.

\clearpage
\section{Measurement and analysis of SiV$^-$ time-resolved PL}
\subsection{Methods}
The $\sivn$ are exposed to the continuous-wave 532-nm probe (varying powers) and the DUV laser (5\,Hz) until the system reaches the quasi-equilibrium (see Sec.\ref{sec:pulsed-pump-probe-measurements}).
Each DUV pulse is accompanied by an RF pulse used to trigger the start of a detection window during which arrival times are recorded.
The resulting set of detection times are binned into either $10^3$ or $10^5$ windows of 200- or 2-\textmu s resolution respectively.

\subsection{Fit parameters}
\begin{figure}[h]
        \centering
        \includegraphics{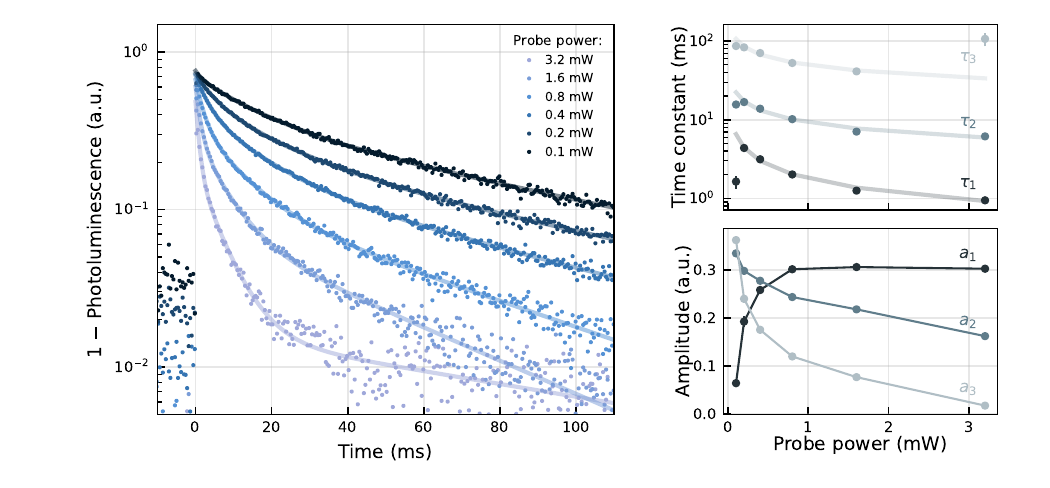}
        \caption{\textbf{Fit of time-resolved SiV$^-$ PL.}
            Time-resolved $\sivn$ PL and corresponding fits to the triple-exponential model.
            The three time constants (top right) approximately follow inverse power laws scaling as $p^{-0.5}$ for $\tau_1$ and $p^{-0.3}$ for $\tau_2$ and $\tau_3$ respectively.
            The normalized amplitudes $a_j$ are shown on the bottom right.
        }
        \label{fig:recovery_fit}
    \end{figure}

The $10^3$-bin dataset is fit to a triple exponential
\begin{equation}
    I(t) = a_0 \big(1 - a_1 e^{-t/\tau_1} - a_2 e^{-t/\tau_2} - a_3 e^{-t/\tau_3} \big)
\end{equation}
where $\tau_j$ and $\alpha_j$ ($j=1,2,3$) are the time constants and associated amplitudes.
The parameter $\alpha_0$ accounts for the difference in measured intensity (due to power and integration time).
Fig.~\ref{fig:recovery_fit} shows the PL signal, recast as one minus the PL, i.e. $(1 - I(t)/a_0)$, to assist in readability of the fit accuracy and data.
The observed time constants span the full measurable range of 1--100\,ms and is shown on the top right.
The time constants roughly follow an inverse power law, but do not appear to scale either linearly, nor quadratically, indicating that they are likely results of complex processes.
We find that the two slower-rates $\tau_2$ and $\tau_3$ decrease in amplitude at larger powers whereas the fast rate $\tau_1$ increases in amplitude.

\clearpage
\section{Analysis of SiV$^0$ spectra}
\subsection{Pre-processing and analysis}
\begin{figure}[h]
    \centering
    \includegraphics{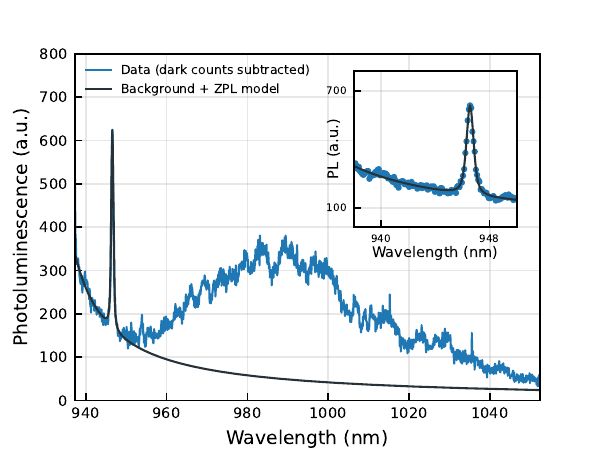}
    \caption{\textbf{SiV$^0$ background model.}
        $\sivz$ PL data with fitted ZPL and background model.
        The fitted region of the curve is shown in the inset ranging from 938--950\,nm.
        Values of ZPL intensity are extracted from the fit to the Voigt lineshape.
    }
    \label{fig:siv0_background}
\end{figure}

The $\sivz$ spectra lies on a downward-sloping background signal which is believed to be the tail of the first-order Raman line of the 830-nm excitation (centered at approximately 935\,nm).
We remove the background by fitting the 946-nm ZPL (and surrounding region below the 951-nm ZPL) to the model
\begin{equation}
    I(\lambda) = aV(\lambda,\sigma,\Gamma) + \frac{b_0}{\lambda - b_1}
\end{equation}
where $V(\lambda,\sigma,\Gamma)$ is a Voigt lineshape (normalized) with parameters $\sigma$ and $\Gamma$ and $(a,b_0,b_1)$ are fit parameters.
The raw spectra are plotted alongside the fitted background in Fig.~\ref{fig:siv0_background}.
Utilize the fit parameter $a$ to characterize the intensity of the $\sivz$ PL (as the large PSB is susceptible to error of the background model).
After subtracting the background component $b_0/(\lambda-b_1)$, we then perform similar outlier removal as for the room-temperature NV data.

\subsection{Stabilty of SiV$^0$}
\begin{figure}[h]
    \centering
    \includegraphics{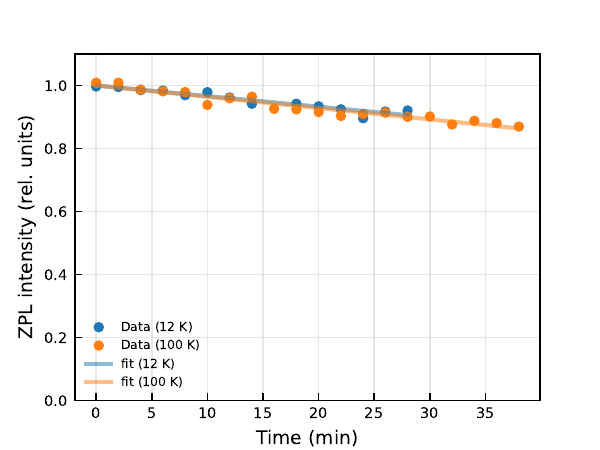}
    \caption{\textbf{SiV$^0$ stability under 830-nm excitation.}
        Normalized amplitude of the ZPL as a function of time under 5\,mW of 830-nm excitation at 12 and 100\,K.
        The signal diminishes by about 10\% after 30 min indicating significant robustness to high-power off-resonant excitation.
        Note that some, if not all, of the reduction can be associated with thermal drift of the sample focus over the course of the measurement.
    }
    \label{fig:siv_stability}
\end{figure}

We performed long series of repeated PL spectra on the $\sivz$ (after one minute of 5-Hz DUV exposure) at both 12 and 100\,K using approximately 4\,mW of 830-nm CW laser.
The fitted peaks are then used to determine the effect of extended 830-nm exposure on $\sivz$ as plotted in Fig.~\ref{fig:siv_stability}.
Over about 30 minutes of continual exposure we observe about a $10$\% reduction of the $\sivz$ 946\,nm ZPL intensity.
The lack of any substantial reduction of PL starkly contrasts the effect of the 532-nm laser which, for an identically prepared sample, would reduce dramatically within a few seconds as shown in the main text Figs.~3--5.
The observed reduction may not necessarily be attributed to the 830-nm excitation itself, but could result from slow drift of the microscope position and focus over the course of the experiment.
These measurements indicate that $\sivz$ is extremely charge stable under off-resonant excitation, indicating its potential for use in quantum technologies.

\clearpage
\bibliography{supplement}